# Into the Blue: AO Science with MagAO in the Visible

Laird M. Close [1a], Jared R. Males[a], Katherine B. Follette[a], Phil Hinz[a], Katie Morzinski[a], Ya-Lin Wu[a], Derek Kopon[d], Armando Riccardi[c], Simone Esposito[c], Alfio Puglisi[c], Enrico Pinna[c], Marco Xompero[c], Runa Briguglio[c], Fernando Quiros-Pacheco[c]

[a]CAAO, Steward Observatory, University of Arizona, Tucson AZ 85721, USA;
[b]OCIW, 813 Santa Barbara St. Pasadena, CA 91101, USA
[c]INAF-Osservatorio Astrofisico di Arcetri, Largo E. Fermi 5, I-50125 Firenze, Italy
[d]MPIA, Germany

## ABSTRACT

We review astronomical results in the visible ($\lambda < 1\mu m$) with adaptive optics. Other than a brief period in the early 1990s, there has been little astronomical science done in the visible with AO until recently. The most productive visible AO system to date is our 6.5m Magellan telescope AO system (MagAO). MagAO is an advanced Adaptive Secondary system at the Magellan 6.5m in Chile. This secondary has 585 actuators with < 1 msec response times (0.7 ms typically). We use a pyramid wavefront sensor. The relatively small actuator pitch (~23 cm/subap) allows moderate Strehls to be obtained in the visible (0.63-1.05 microns). We use a CCD AO science camera called "VisAO". On-sky long exposures (60s) achieve <30mas resolutions, 30% Strehls at 0.62 microns (r') with the VisAO camera in 0.5" seeing with bright R < 8 mag stars. These relatively high visible wavelength Strehls are made possible by our powerful combination of a next generation ASM and a Pyramid WFS with 378 controlled modes and 1000 Hz loop frequency. We'll review the key steps to having good performance in the visible and review the exciting new AO visible science opportunities and refereed publications in both broad-band (r,i,z,Y) and at Halpha for exoplanets, protoplanetary disks, young stars, and emission line jets. These examples highlight the power of visible AO to probe circumstellar regions/spatial resolutions that would otherwise require much larger diameter telescopes with classical infrared AO cameras.

# 1.0 INTRODUCTION

## 1.1 Review of AO Astronomy in the Visible

In the early years of AO (1990-1994) AO work (often by the military) was done in the visible. The 1.5m Starfire AO telescope started with visible cameras. Strehls were quite low ~10% at 0.85µm, and resolutions were also fairly low at ~0.21-0.25" FWHM (Drummond et al. 1994). The overall use of AO for astronomical science (at any wavelength) was still fairly rare. Moreover, once HST was repaired in late 1993 visible AO observations lost favor with astronomers who switched to using the newly available NIR (NICMOS/InSb) arrays for AO astronomical observations. In this manner there was no direct competition from HST superior performance in the visible.

---

[1] lclose@as.arizona.edu; phone +1 520 626 5992

In the later years (1994-2012) the 1-2.5μm wavelength range (the NIR) proved a great scientific home for AO astronomical science (still mostly independent of direct HST competition) and produced Strehls that were high (~60% at K) on 4-8m class telescopes. Visible AO had little significant astronomical science impact in astronomy. But the visible AO was used for Solar, Military (AEOS/VisIM), and Amateur AO (often with some form of Lucky imaging). Nevertheless, there were almost no astronomical refereed papers in the visible from 1994-2012. The most likely reasons for this are: 1) the Strehls were not very good in the visible (often <1%) with only ~100 corrected modes; and 2) all the visible photons are typically used for the wavefront sensor. Visible AO science really required that an AO system be designed from the start to work in the visible.

### 1.3 Scientific Advantages to Visible AO

Visible has many scientific advantages over the NIR. After all, most astronomy is done in the visible, but almost no AO science is done with $\lambda<1\mu m$ on large 6.5-10m class telescopes. A short list of some of the advantages of AO science in the visible compared to the NIR is:

-- **Better science detectors** (CCDs): much lower dark current, lower readnoise, much better cosmetics (no bad pixels), ~40x more linear, and camera optics can be warm and simple.
-- **Much Darker skies:** the visible sky is 100-10,000x darker than the K-band sky.
-- **Strong Emission lines**: access to the primary recombination lines of Hydrogen (Hα 0.6563 μm) --- most of the strongest emission lines are all in the visible
-- **Off the Rayleigh Jeans tail**: Sun-like stars have much greater range of colors in the visible (wider range color mag diagrams) compared the NIR.
-- **Higher spatial resolution**: The 20 mas resolution regime opens up. An visible AO system at r band ($\lambda=0.62\mu m$) on a 6.5m telescope has the spatial resolution (~20 mas) that would otherwise require a 23m ELT (like the Giant Magellan Telescope) in the K-band. So visible AO can produce ELT like NIR resolutions on today's 6.5-8m class telescopes.

### 1.4 Keys to good AO Performance in the Visible

While it is certainly clear that there are great advantages to doing AO science in the visible it is also true that there are real challenges to getting visible AO to work with even moderate Strehls on large telescopes. Below we outline (in rough order of importance) the most basic requirements to have a visible AO system on a 6.5m sized telescope.

1. **Good 0.6" Seeing Site** – Large $r_o$ (long $\tau_o$) and consistency (like clear weather, low humidity) is critical.
2. **Good DM and fast non-aliasing WFS**: need many (>500) actuators (with $d<r_{o/2}$ sampling), no "bad" actuators, need at least >200 well corrected modes.
3. **Minimize all non-common path (NCP) Errors:** Stiff "piggyback" design with simple Visible science camera well coupled to the WFS –keep complex optics (like the ADC) on the common path. Keep optical design simple and common as possible.
4. **Lab Testing:** Lots (and lots) of "end-to-end" closed loop testing with visible science camera.
5. **Modeling/Design:** Well understood error budget feeding into analytical models, must at least expect ~135 nm rms WFE on-sky. Try to measure/eliminate vibrations from the scope.
6 **High Quality Interaction Matrixes:** Excellent on-telescope IMATs with final/on-sky pupil. Use close-loop to increase the SNR of the high order modes.
7 **IR camera simultaneous with Visible AO camera:** this is important since you achieve a 200% efficiency boost. It also allows for excellent (pain free) contingency in poor seeing when only NIR science is possible.

## 1.5 Visible AO vs. HST, Interferometers, or NIR AO on 8-10m Telescopes

The Magellan AO system (MagAO) is a good example of a visible AO system. MagAO can regularly obtain (>50% of the nights so far observed) moderate Strehl (~20% at 0.65μm) and 20-30 mas resolution images in the visible (Close et al. 2013; Fig. 1). *This is >2x higher resolution than the Hubble Space Telescope can achieve at the same wavelengths, and is also ~2-3x better than the sharpest images one can make from the ground with conventional NIR AO on the largest 8-10m apertures.* While interferometers can provide higher spatial resolutions, their limited "uv" coverage, limiting magnitudes, and very small FOV (<0.1") make them generally much less attractive than direct imaging for the science cases outlined in section 2.3. Also, speckle interferometry can achieve the diffraction-limit, but is only effective on the brightest binary stars in the optical, with limited contrast and, hence, dynamic range. Whereas, MagAO has detected exoplanets $10^{-5}$ times fainter within 0.5" of the host star (Males et al. 2014) such contrasts are impossible to achieve with any speckle or interferometric techniques.

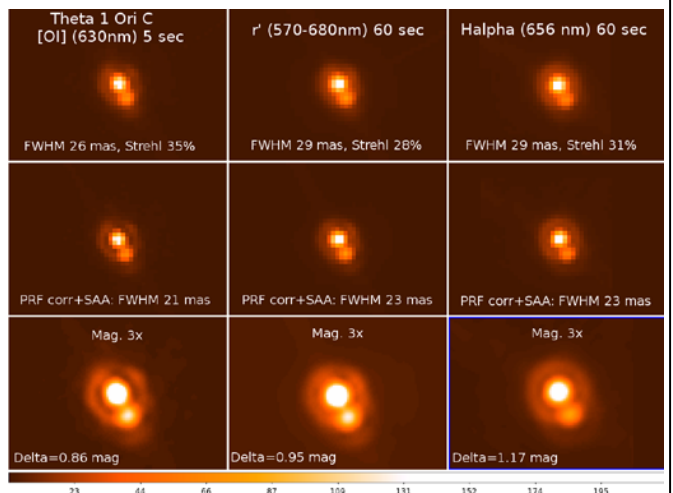

*Fig 1: Top row*: the central ionizing binary of the Trapezium: $\theta^1$ Ori C as imaged with MagAO's Visible CCD camera (VisAO) in different filters. Note the excellent resolution in the raw 60 second image. We note that no post-detection shift and add (SAA) was applied, nor was there any frame selection used to produce these top row images. Typically we achieved resolutions of 0.026-0.029" and Strehls of 28-35% in 0.5-0.7" V-band seeing. *Middle row:* same as top row, except the images have been post-detection aligned (SAA) and the pixel response function (PRF) has been removed. This improved image resolution by~5-6 mas. *Bottom row:* is just magnified by 3x These are the highest resolution, deep, images ever published to our knowledge (modified from Close et al. 2013).

A simple reason that many "VisAO" science cases *cannot* be done with *HST* is that the brightest science targets (V<$8^{th}$ mag) become difficult to observe without debilitating core saturation/charge bleeding -- even for minimum exposures. Moreover, the permanent loss of the ACS HRC channel leaves just the visible coronagraphic "wedge" in STIS on *HST*. With a size of >0.2", which often covers up the most important science area for circumstellar science (the core), this bar inhibits HST study of our "VisAO" science cases. Also we note that *JWST* will not likely produce diffraction-limited imaging in the visible. **MagAO with its VisAO camera provides a 2-3 fold improvement in the angular resolution of direct imaging in astronomy while simultaneously gaining access to the important narrowband visible (0.6-1.05 μm) spectroscopic features (like the key hydrogen recombination Halpha line at 0.656 μm) that have been inaccessible at 0.02" resolutions to date (Close et al. 2013; Close et al. 2014).**

Visible AO is being used on smaller telescopes still today. An excellent example of this is the Robo-AO system in use at the 1.5m at Palomar. It uses an UV LGS to allow remote scanning of the skies. Robo-AO can do thousands of AO targets very quickly. Although the spatial resolutions are limited ~0.2-0.3" the ability to image thousands of targets quickly is very powerful. For more on Robo-AO see Baranec et al. and Law et al. these proceedings. Robo-AO's follow-up survey of thousands of *Kepler* field KOI's is an excellent example of visible AO's scientific power.

There are other visible AO instruments on larger telescopes like those that are fed by the very high order PALM-3000 AO system at the 5m Hale at Palomar. Examples of its visible science cameras include the SWIFT IFU and TMAS CCD camera which has made excellent images of the moons of Jupiter (Dekany et al. 2012; Dekany et al. these proceedings). There is also the VLT's SPHERE ZIMPOL camera which has just had first light (Bezuit et al. these proceedings). Although none of these visible cameras have yet produced refereed visible astronomic science results --they will likely do so in the very near future. However, at the time of these proceedings, the MagAO/VisAO system has produced significant visible AO astronomical science (in seven different refereed publications) as is described later in this paper.

There are also more visible AO cameras coming on-line soon. The VLT MUSE IFU has a GLAO feed from the DSM of the ~2016 era AOF and an optional diffraction-limited optional platescale. In the far future Keck's proposed (but not yet fully funded) NGAO could have higher resolution (and LGSs leading to wide sky coverage). It also true that the future of *HST* is in some doubt and once *HST* stops scientific operations (possibility at any time now) the only access astronomers will have to diffraction-limit visible images will be through visible AO. So this will be an increasingly important field for AO science and astronomy in general. So there will be a long-term need for VisAO science with MagAO and other visible AO systems for the foreseeable future.

# 2.0 INTRODUCTION TO MagAO

To best understand how the MagAO system works in the visible first need to understand a bit about the history of the Adaptive Secondary Mirror (ASM).

## 2.1 Past Developments of Adaptive Secondary Mirrors for Adaptive Optics

Adaptive secondary mirrors (ASMs) have several advantages over conventional deformable mirrors: 1) they add no extra warm optical surfaces to the telescope (so throughput and emissivity are optimal); 2) the large size of the optic allows for a relatively large number of actuators and a large stroke; 3) their large size enables a wide (>5') field of view (FOV); 4) the non-contact voice-coil actuator eliminates DM print-through; and hence 5) performance loss is minor even if up to ~10% of the actuators are disabled (proof: all data in this paper was obtained with ~12 of 585 MagAO actuators disabled). They also give better "on-sky" correction than any other AO DM (see Fig. 1). Hence, adaptive secondaries are a transformational AO technology that can lead to powerful new science and telescopic advancement (Lloyd-Hart 2000). MagAO is the result of 20 years of development by Steward Observatory and our research partner INAF-Osservatorio Astrofisico di Arcetri of Italy and industrial partners Microgate and ADS of Italy.

In 2002 this Arizona/Italy partnership (Wildi et al. 2002) equipped the 6.5m MMT with the world's first ASM. This ASM is a 65 cm aspheric convex hyperboloid Zerodur shell 2.0 mm thick. The thin shell has 336 magnets bonded to its backside where 336 voice-coil actuators with capacitive sensors can set the shell position. The MMTAO has carried out regular NIR science observations since 2003 reliably with little down time (see for example: Close et al. 2003a; Kenworthy et al. 2004, 2007, 2009, Heinze et al. 2010; Hinz et al. 2010). However, the MMT system was really a prototype ASM.

From the many lessons learned from the MMT's ASM, a new "2$^{nd}$ generation" of ASMs were fabricated for the LBT and MagAO. LBT's AO system has had a spectacular on-sky first light in June 2010 (Esposito et al. 2010) obtaining the best AO performance of a large telescope to date. ESO has also developed (a larger 1.1m, thicker ~2.0 mm) DSM shell for their future AO facility for science use in ~2015 (AOF; see Arsenault et al. 2014). Also such ASMs are baselined for the 24m Giant Magellan

Telescope (GMT) secondaries (~2024) and the M4 of the ~39m E-ELT (~2024?), and perhaps as a future upgrade to the secondary of the 30m TMT. *Adaptive secondaries are now key to 3 major AO systems and will likely play a role in all future large telescope projects.*

## 2.2 The 2nd Generation 585 element ASM for MagAO

Our "thin shell/voice coil/capacitive sensor" architecture is the only proven ASM approach. MagAO's "LBT-style" 2nd generation 585 actuator 85cm dia. ASM offers many improvements over the 1st generation "MMT" ASM. In particular, MagAO's successful Electro Mechanical acceptance tests in June 2010 proved that the MagAO 585 ASM has larger stroke (±15 µm), a thinner shell (at 1.6 mm vs. 2.0 mm), half the "go to" time (<0.7-1.0ms; with electronic damping), 2-5 nm rms of positional accuracy (by use of a 70 kHz capacitive closed-loop), and just 30 nm rms of residual optical static polishing errors (compared to ~100 nm rms on the current MMT shell). These improvements are taken advantage of by LBTAO as well (Riccardi et al. 2010), but MagAO's mirror overall is slightly better behaved compared to LBT (MagAO's lack of 87 slower "outer ring" LBT actuators increases its speed w.r.t LBT). Moreover, MagAO's ASM is much more flexible than any other ASM, while also not having the inner "stressed" hole illuminated (due to the 0.29 central obscuration of Magellan). So it is not really surprising that MagAO should be the highest performance ASM yet built – see Fig. 1 for proof of how effective MagAO is at high-order correction.

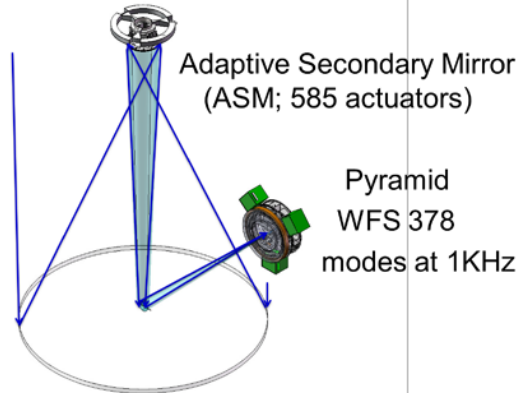
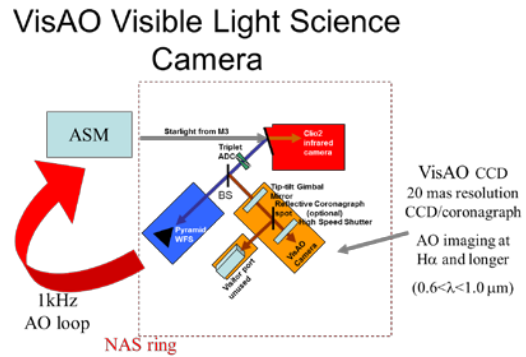

*Fig 2:* Schematic diagrams of the MagAO systems and VisAO camera.

Magellan is a Gregorian telescope (Fig. 1a) and requires a large (d=85.1 cm) concave ellipsoidal ASM The concave shape of a Gregorian secondary enables easy testing "off the sky" with an artificial "star" for daytime tests in the dome. In addition, 585 mode servo loop CPU latency is limited to <120 µs through the use of 132 dedicated DSPs (producing ~250 Giga Flops, in the ASM electronics) for very fast real-time performance.

## 2.3 Current Status of the MagAO System

MagAO had first light (with its science cameras VisAO and Clio2; see figure 1) in November 2012 (as scheduled after the PDR that occurred in 2009). During first light the AO performance was excellent in the visible with just 200-250 modes in closed loop (see Close et al. 2013). The system than had a major upgrade in the second commissioning run in April 2013 when the maximum number of corrected modes was raised to 378 and corrections as good as 102 nm rms were obtained on bright stars in median 0.6" seeing. Then from April 1 to 25th (in the 2014A semester of Magellan) the first open facility science run was executed. The demand for MagAO observing time in 2014A was very high (typically 3-5x oversubscribed). The individual TACs from Carnegie, MIT, Harvard/CfA, Arizona, University of

Michigan, and Australia all assigned time for their users on MagAO in 2014A. The 2014A run was a good success with almost all science programs executed and with less than 2% downtime from MagAO and its cameras. Overall the community was very happy with MagAO's performance. In 2014B these TACs have now assigned 37 nights to MagAO in the November-December timeframe of 2014. MagAO is the facility AO system for the Magellan community. See Morzinski et al. 2014 (these proceedings) for more details about MagAO and its infrared camera Clio2.

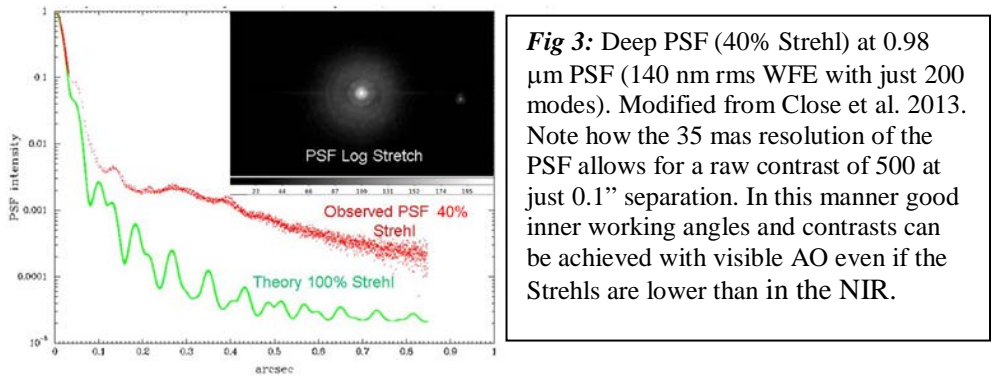

*Fig 3:* Deep PSF (40% Strehl) at 0.98 μm PSF (140 nm rms WFE with just 200 modes). Modified from Close et al. 2013. Note how the 35 mas resolution of the PSF allows for a raw contrast of 500 at just 0.1" separation. In this manner good inner working angles and contrasts can be achieved with visible AO even if the Strehls are lower than in the NIR.

The schematic drawing above (Fig. 2b) outlines how our VisAO and Clio2 cameras are co-mounted and can be used simultaneously (if desired). No instrument changes are ever needed to switch between IR and visible science. In the campaign/queue mode envisioned for MagAO one will not use VisAO if seeing is poor, nor will any telescope time be lost since 1-5.3 μm Clio2 science can be done in >90% of the seeing conditions at Las Campanas --an excellent (median V=0.64" ) seeing site (Thomas-Osip 2008).

**2.4 Our Simulated MagAO/VisAO Error Budget Compared to On-Sky results**
MagAO's 585 controllable modes map to a 23 cm "pitch" on the 6.5m primary. This is a smaller (tighter) pitch than all current AO systems (even 20% smaller than the LBT). To predict the exact degree of correction we used "end-to-end" simulation of MagAO/VisAO with the Code for Adaptive Optics Simulation (CAOS; Carbillet et al. 2005). Our CAOS simulations (assuming no extra telescope vibration) predict slightly larger wavefront errors (135 vs. 122 nm rms) than our "corrected" test tower results (Males et al. 2012; Close et al. 2012), hence our test tower results are quite consistent with our analytical model of MagAO.

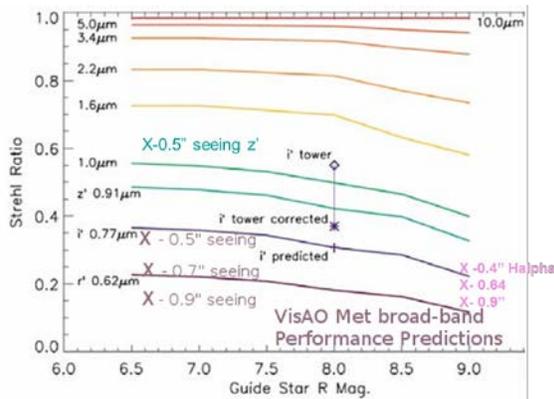

*Fig 4:* MagAO has achieved 104-140 nm of wavefront error using 200-378 modes in 0.65" seeing (median conditions) on sky. This performance is very similar to the predictions by Males et al. 2013 for the MagAO. The curves are the predicted correction and the "X" are a few observed Strehls color coded to match the corresponding curve. See Close et al. 2014, fig 2 for 3,256 more data points with 19-21% Strehls at 0.65μm in 0.4-0.8" seeing.

*2.4.1 Moderate Strehls robustly obtained in the Visible with MagAO on-sky*

While simulations, and lab tests give significant comfort, the true test is whether the AO system can actually achieve >30% Strehls reliably in median atmospheric conditions (0.7") on the sky. Since fall 2012 MagAO has been tested on-sky. Figure 4 shows that the published Strehls achieved by MagAO/VisAO are right along the expected values. We now have 3 MagAO runs with diffraction-limited images obtained with VisAO on most of the nights. Diffraction-limited visible imaging with MagAO/VisAO is almost guaranteed at Magellan in median seeing/wind conditions on V<11 mag guide stars (Close et al. 2014).

**2.3 A Few Selected Science Cases for Visible AO Imaging (MagAO/VisAO) Observations**

*2.3.1 Imaging of Young Stars and Proplyds in Orion*

Clearly the exciting possibility of obtaining ~20 mas FWHM images with MagAO could enhance our understanding of the positions (and motions) of the nearest massive young stars. Hence we targeted the Orion Trapezium cluster during the first light commissioning run with the MagAO system. Indeed in Figure 1 we can see how well the 32 mas binary Theta 1 Ori is split with visible AO (Close et al. 2013). As we can see in figure 5 in the trapezium region alone many scientific results can be obtained with visible AO. We can measure the astrometry of the trapezium members down to 0.2 mas/yr which is excellent (Close et al. 2013). This translates to velocities of 0.4 km/s, at which point the individual orbits start to show orbital arcs in the Trapezium (Close et al. 2013).

As figure 5 also illustrates we can image off-axis proplyds at high resolution. These warped protoplanetary disks are being photoevaporated by the intense UV radiation and solar wind from theta 1 Ori C. Interestingly Wu et al. 2013 were able to use it as a guide star to make HST resolution images of LV1 6.3" off-axis from the guide star. This suggests (as do the ~10" corrected FOV from SCAO solar AO systems) that the corrected FOV in the optical can be (even at Halpha) as large as 6".

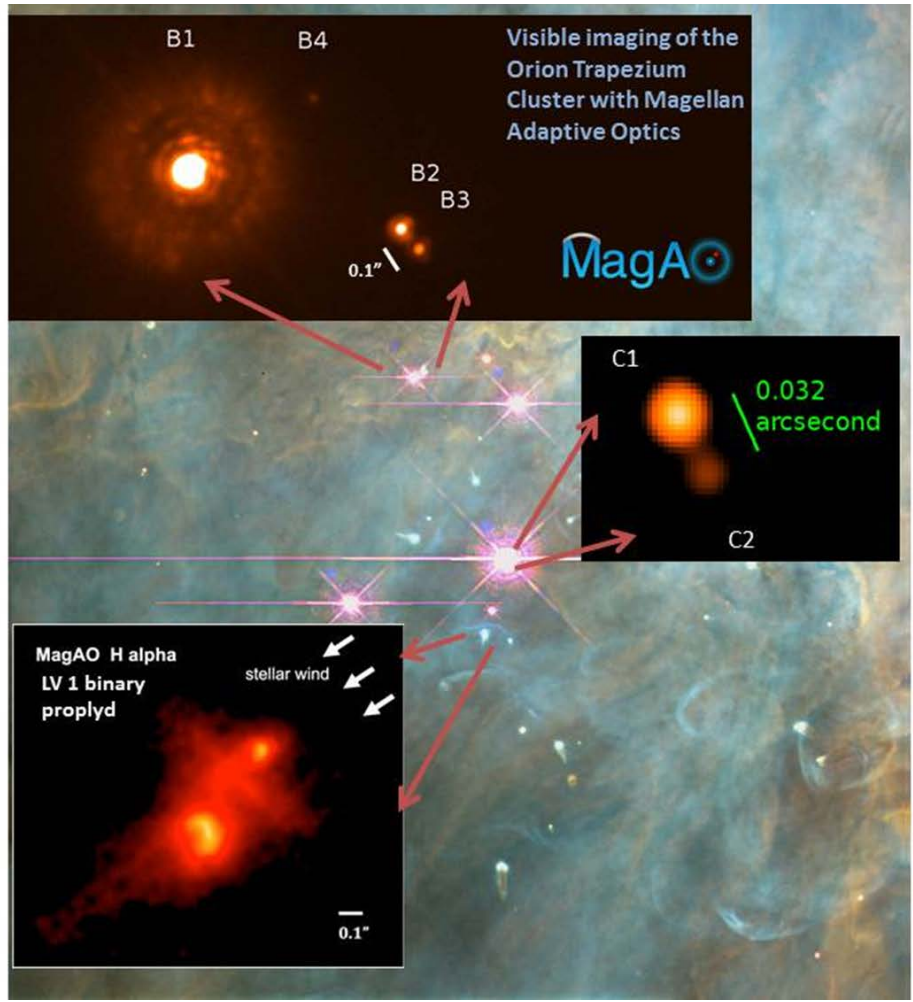

*Fig 5:* Best images of the young stars (and their orbits) in the famous Trapezium cluster in the Orion Nebula. Close et al. 2013 and Wu et al. 2013. Background image from NASA/*HST* archive.

### 2.3.2 High-Contrast Imaging with VisAO

Another Key science field for visible AO is high contrast imaging of exoplanets with direct imaging. The large ~10 Jupiter mass planet Beta Pic b some ~9 AU from its host star is an excellent target for this sort of work. At the first light of MagAO deep imaging of Beta Pic was carried out to see how faint the planet would appear in the Ys band. The full results of this first detection of an exoplanet with a CCD from the ground can be found in Males et al. (2014). In figure 6 we show what the contrast curve was from 2.5 hours of open shutter observation of the system. We note that this achieved contrast is very similar to that obtained by GPI on Beta-Pic in 30min in H band with SSDI/TLOCI (http://www.gemini.edu/sciops/instruments/gpi/instrument-performance?q=node/11552 ). The important fact here is that the smaller $\lambda/d$ of visible AO can compensate for poorer (40%) Strehls to enable similar contrasts of an "extreme AO" system like GPI working at H band.

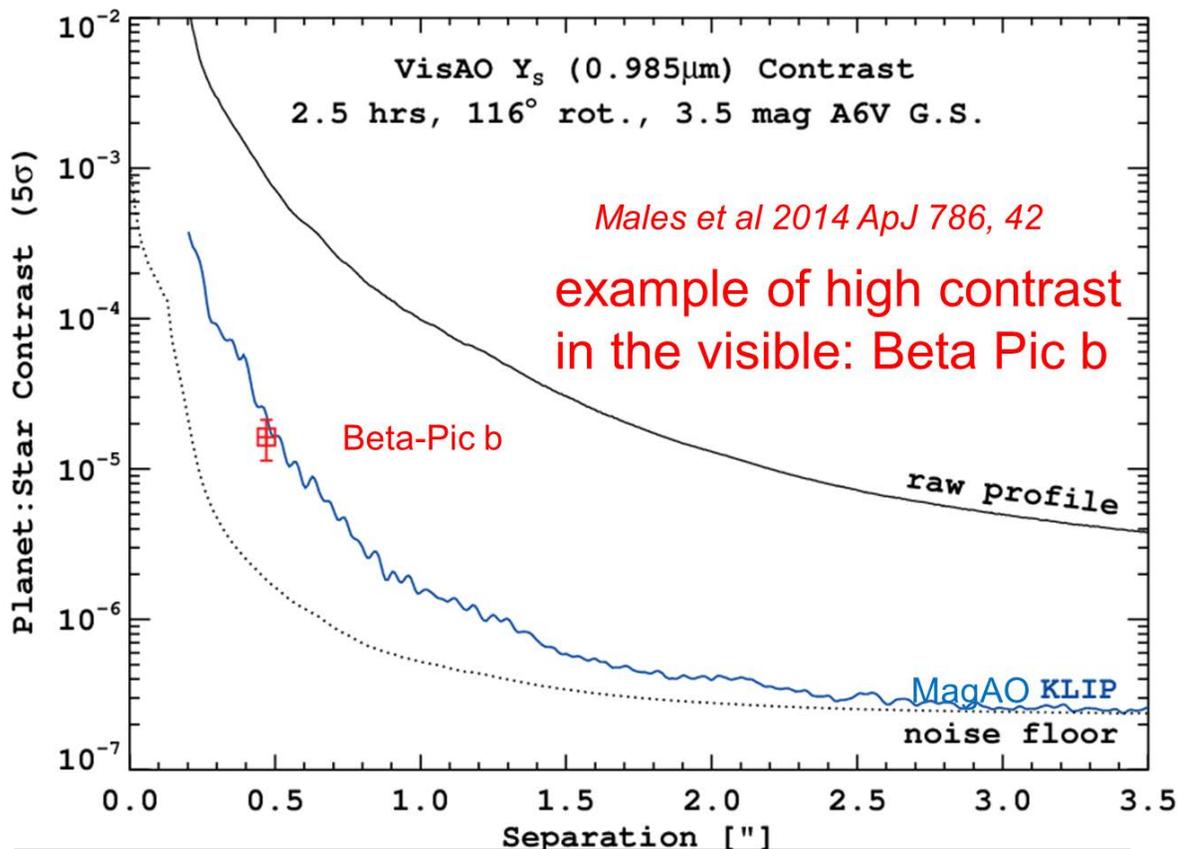

*Fig 6:* This "first light" visible AO contrast plot for MagAO from the published results of Males et al. (2014). It shows a similar contrast as that of the Gemini Planet Imager (GPI) at H band in 30 min (also on Beta-Pic). Visible AO is also "extreme AO" in terms of contrast.

*2.3.3 High Contrast imaging with a novel SDI Camera*
In the last section we learned that visible AO can make high contrast images in the visible broad band with ADI/PCA reductions. Indeed the VisAO camera design incorporates all the key features and remotely-selectable elements necessary to optimize such visible AO science. In particular, the coronagraph wheel contains a range of our custom reflective ND masks (Park et al. 2007), allowing deep circumstellar science on bright targets that would otherwise saturate the detector. The main purpose of these masks is to prevent blooming of the CCD47 VisAO detector. Our coronagraph doesn't need to suppress diffraction rings since the ADI/PCA reduction technique work very well. However the number of useful astrophysical investigations possible with Strehls in the ~20-40% range are limited without accurate PSF calibration. In the next section we introduce a new "SDI" mode of the VisAO camera to calibrate the PSF directly.

One needs simultaneous PSF information to compare to (or deconvolve against) the "in-line" science image. An extremely effective technique for this is Simultaneous Differential Imaging (SDI; Close et al. 2005) which utilizes a Wollaston Beamsplitter to obtain nearly identical, simultaneous, images of the *o-polarized* and *e-polarized* PSFs (typically there is <10 nm rms of non-common path SDI error between the *o* and *e* images; Lensen et al. 2004; Close et al 2004). The SDI configuration of the VisAO camera includes: 1) a thin small angle calcite Wollaston beamsplitter near our pupil; and 2) a split on-Hα/off Hα "SDI" filter just before the focal plane (for the *e* beam/*o* beam; see fig. 7). In this mode, we have obtained an almost perfect (photon noise limited) simultaneous calibration of the PSF "off" and "on" the Hα line on-sky. Hence, a simple subtraction of the "off" image from the "on" image will map Hα structures (jets, disks, accreting faint companions etc.), with minimal confusion from the continuum or PSF. We have three such SDI filter sets (with optional SDI double spot coronagraphic masks) for the Hα, [OI], and [SII]. Our new (as of April 2014) Halpha SDI filters yield on-sky 1 hr. images of $2\times10^{-6}$ contrast at 1" which exactly at the photon-noise limited floor.

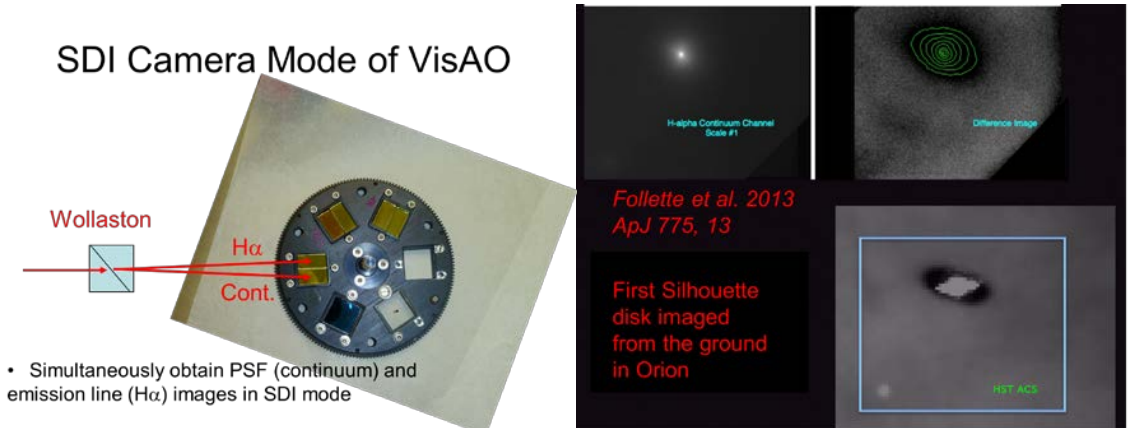

*Fig 7:* A cartoon of the SDI mode (left). To the right we see a real science example of the SDI mode in action. Here we see how a young forming solar system has its dust traced out by the shadow it casts as the bright Halpha background is absorbed by its dust disk. Modified from Follette et al. 2013.

*2.3.4 High-Contrast Hα Emission Line Imaging of Young Extrasolar Planets: GAPlanetsS Survey*
Another key survey project we are doing with VisAO is high-contrast Hα SDI imaging of the emission from the accretion shock caused by gas accreting onto gas giants during formation. We'll target >30 known southern, young (<10 Myr), nearby (<150pc), gas-rich transitional disk systems that have I≤10 mag. During the time of gas accretion the protoplanets will have $\sim10^{-(3-4)}$ $L_{sun}$ and much of this will be radiated at Hα for $\sim10^6$ yrs (Fortney et al. 2008). In this manner, we can finally directly image giant planets where/when they form (likely past the "snow-line" >30-50 mas) as Hα point-sources orbiting the young target star. This GAPplanetS survey is already 60% complete, and will be likely finished after the 2014B run this fall (Follette et al. 2014).

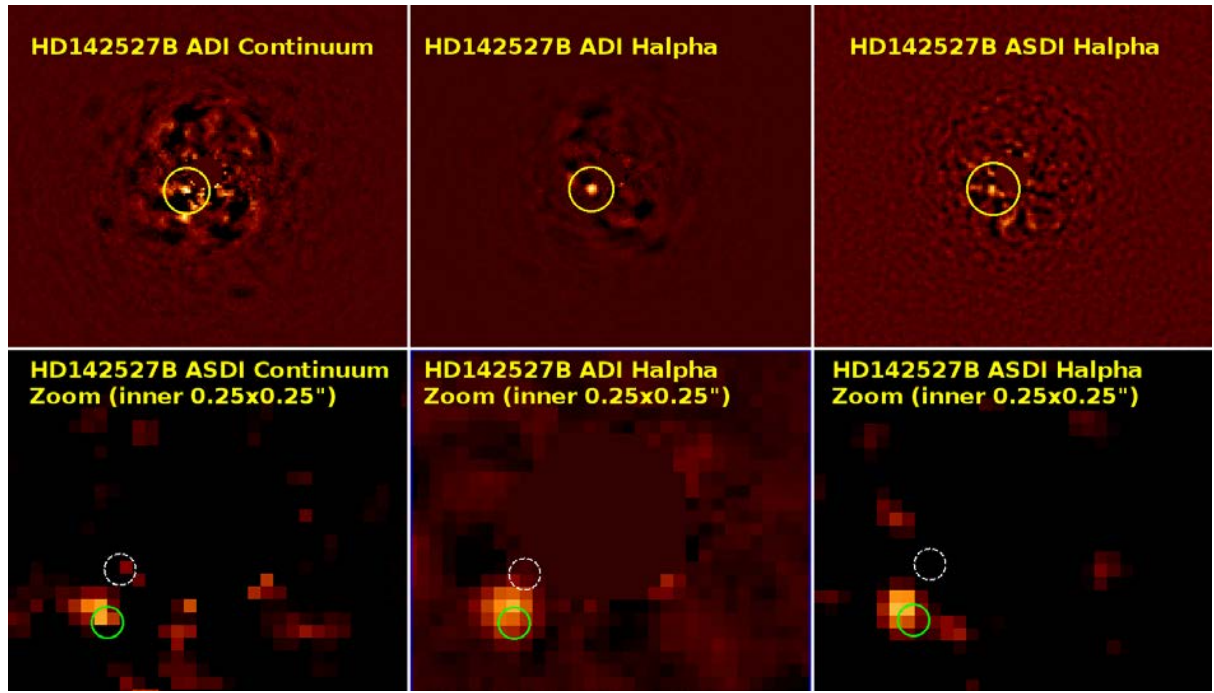

**Fig 8:** Here we examine the first GAPplanetS target, the famous transitional disk HD142527 with the SDI mode. (*left*) Continuum (643 nm) ADI reduced image. (*Bottom Zooms*): Note the weak (~3 sigma) detection near the location of the candidate of Biller et al. 2012 (green circle). The source is inconsistent with the background star position (white circle). (*middle*) Hα ADI images. Note the unambiguous 10.5 sigma Hα point-source at sep=86.3 mas, PA=126.6°, hereafter HD142527B. (*right*) ASDI data reduction, here NCP narrow-band filter ghosts are not as well removed as with ADI. NOTE: the new single substrate SDI filters designed by custom scientific (and installed in VisAO in April 2014) has removed all signs of filter ghosts in the current Halpha SDI camera.

These images clearly detect an accreting young companion just 86 mas (and 1000x fainter in continuum) from the primary star! Modified from Close et al. (2014).

*2.3.5 High-Contrast Imaging of Extrasolar Planets in the Habitable Zone of Nearby Planets*
An other Key Project for the VisAO camera will be to take deep ADI datasets on nearby stars to directly detect reflected light from giant planets. Jared Males has formed a strong group of scientific collaborators from across the Magellan Partnership to make sure that such a survey is well designed and has been obtaining the telescope resources required to finish. Please see Males et al. in these proceedings for more details.

*2.3.6 Asteroid and Solar System Surfaces, and Titan's Atmosphere*
There are many science cases where simply mapping objects at these very high resolutions is exciting. For example, even moderate Strehls are fine for mapping the edges of astroidal surfaces with the VisAO imager. The VisAO camera can also map Titan (diameter ~0.7") with a 0.95μm $CH_4$ filter. Similar work is being done with the Palm-3000 system, where the moons of Juipter and Saturn are being imaged in the optical (Hildebrandt, S. private comm.)

# 3.0 CONCLUSIONS

  The future of visible AO is clearly very bright. We are now in a new age where AO correction is good enough for excellent visible AO science to be carried out with facility AO systems like MagAO. The near future will bring many new visible AO systems on-line and many more exciting science results will follow. We already have exciting MagAO/VisAO science results on exoplanets (Males et al. 2014; Bailey et al. 2014; Close et al. 2014; Morzinski et al. 2014; Follette et al. 2014; Biller et al. 2014), young stars (Close et al. 2013; 2014), circumstellar disks (Follette et al. 2013; Wu et al. 2013; Biller et al. 2014) and Halpha emission jets (Monnier et al. in prep) with many more science papers on these and other fields soon to be published.
  In the future (post-*HST*) the only source of diffraction-limited visible images will be visible AO. Sky coverage is currently small --as visible AO is limited to NGS targets V<12 mag. But the use of LGS with visible AO could increase sky coverage considerably. Making visible AO a replacement for many of *HST's* science programs a viable option.

  The MagAO ASM was developed with support from the NSF MRI program. The MagAO PWFS was developed with help from the NSF TSIP program and the Magellan partners. The VisAO camera and commissioning were supported with help from the NSF ATI program. The ASM was developed with support from the excellent teams at Steward Observatory Mirror Lab/CAAO (University of Arizona), Microgate (Italy), and ADS (Italy). L.M.C.s research was supported by NSF AAG and NASA Origins of Solar Systems grants. J.R.M. and K.M. were supported under contract with the California Institute of Technology (Caltech) funded by NASA through the Sagan Fellowship Program. We also wish to thank the amazing team at Magellan that helped us commission MagAO at LCO.